\documentclass[11pt]{article}
\usepackage{amssymb,amsmath,amsthm}
\usepackage{rotating}
\usepackage{graphicx,comment}
\usepackage{color}

\excludecomment{omitext}
\begin{document}

\begin{center}

{\bf On the  displacement of two immiscible  Oldroyd-B  fluids in  a  Hele-Shaw cell}

Gelu I. Pa\c sa, Simion Stoilow Institute of Mathematics of Romanian Academy,

P.O. BOX 1-764, RO-14700, Bucharest, Romania
\end{center}

{\it Abstract}.  The Saffman-Taylor instability occurs when  { a} Stokes fluid 
is displaced by a less viscous one { in}  a Hele-Shaw cell. 
{ This  model  is useful to study the secondary oil recovery from a 
pororus medum.}
Since  1960,  { polymer solutions}   were  used as displacing fluids;  moreover, 
the oil in  a porous reservoir can often be  considered a  non-Newtonian fluid. 
Motivated by this fact, in this paper we study the linear stability of the 
displacement of two Oldroyd-B  fluids in a rectilinear     Hele-Shaw cell, even 
if  the direct relevance  for the flow through  porous media  is not so evident.  
We get an { approximate} formula of  the growth rate of perturbations, 
which depends on the { difference} of the Weissenberg  numbers of the 
two fluids. 
{ A blow-up of the growth rate appears for a critical value of this 
difference.} 
This singularity is in agreement with { previous} numerical and 
experimental results, already reported by several papers concerning the flow   
of complex fluids in Hele-Shaw cells.

\vspace{1cm}


\section{Introduction}

 A Hele-Shaw cell is a technical device introduced in \cite{HS}, formed by 
two parallel plates separated by a { narrow}  gap.
{ The equations verified by the mean velocities of a Stokes fluid 
in a Hele-Shaw cell  are  similar with the Darcy's  law for flow in a 
porous medium } - {   see ~\cite{BE},  ~\cite{LA}.
The interface between two immiscible  Stokes fluids   in a Hele-Shaw cell is 
unstable when the displacing fluid  is less viscous - see Saffman and Taylor  
~\cite{ST}.

The Hele-Shaw model can be used to study the  secondary oil recovery. Since 
1960, good results were obtained  by using a  polymer solution as "forerunner" 
in the oil recovery process - see  ~\cite{GH} and the references therein. 
The  oil can often be considered  a  non-Newtonian  fluid, so  it is useful 
to study the stability of the  non-Newtonian displacements in Hele-Shaw cells.
 
The non-Newtonian fluids are studied   in a  large number of papers  - see
 ~\cite{BIRD-60}, ~\cite{FOS}, ~\cite{guillope},  ~\cite{MIR-2001}, 
 ~\cite{RE},   ~\cite{schowlater},   ~\cite{truesdell}.  
This fluids exhibit at least two characteristics not present in Newtonian case: 
shear-thinning  and elasticity. 
Two important constitutive models  { exists}  in order to describe 
these effects. 
The Ostwald-de Waele power-law fluid shows shear-thinning but is  inelastic.
The  Oldroyd-B  model exhibits elasticity but not  shear-thinning.


Important results about  the stability of non-Newtonian  displacements  in Hele-Shaw 
cells  were obtained  in \cite{ARO}, \cite{BES}, \cite{FON},  \cite{MACC},
\cite{MAR-B},  \cite{SAD}, \cite{WI}.

A formula of the growth rate of perturbations was obtained in  \cite{WI}, when
 a power-law fluid (with exponent $q$) is displaced   by air in a rectiliniar cell. 
 It is  multiplied by $q^{-1/2}$ 
as compared with the Newtonian case, but there is no qualitative change.

In  \cite{ARO} { is studied}  the flow near an arbitrary corner for any 
power-law fluid. 
The  variational calculus is used in \cite{BES}, to investigate the time-dependent 
injection rate  that minimises the Saffman -Taylor instability, when  shear-thinning 
fluids, slowly gelling or  fast gelling  polymer solutions are displacing.
In   \cite{FON} are studied some methods for minimize the fingering phenomenon when 
an inviscid fluid displaces a  power-law fluid. 
The problem of bubble contraction in a Hele-Shaw cell is studied in \cite{MACC}, when  
the surrounding fluid is of power-law type, related with a  small perturbation of the 
radially  symmetric problem.  
The stability of the displacement of two  power-law fluids  in a radial Hele-Shaw cell 
has been considered  in   \cite{MAR-B}.
The displacement of a high-viscosity  power-law fluid by a low-viscosity Newtonian 
fluid in a radial Hele-Shaw cell  is studied in  \cite{SAD} and a detailed  analysis 
of the  flow is  given, concerning {  the fractal  fingering  patterns. }


{ Numerical results concerning  the displacement of  Oldroyd-B  and Maxwell upper -
convected fluids  by air in rectilinear Hele-Shaw cells are given in ~\cite{mora2009}, 
~\cite{mora2012}},  ~\cite{WI}. A  blow-up  of the growth constant  was 
reported in ~\cite{WI}, when the relaxation (time) constant  is increasing  up to  
a critical value. This  phenomenon  may be related with the fractures observed in 
the  flows of complex fluids  in Hele-Shaw cells  - see  \cite{nase2008},  
\cite{nittman1985},   \cite{zhao1993} and the references  therein. 


In this paper we study the linear instability of the displacement of two Oldroyd-B 
fluids in a rectilinear Hele-Shaw cell. 
We use the same depth-averaged interface conditions as in ~\cite{ST}, ~\cite{WI}. 
The  thickness of the narrow gap  is much smaller compared with the cell lenght, 
then we can neglect  some terms  in the perturbations  equations. The new element 
is the explicit formula \eqref{COMPAT-10} of the growth rate of perturbations, 
obtained in terms of $(C_i-D_i)$, where $C_i, D_i$ are  the Weissenberg numbers of 
the  two fluids. Most numerical methods fail when the Weisenberg numbers are near 1.  
A blow-up of our growth rate formula \eqref{COMPAT-10}  appears for some  critical   
values   $(C_i-D_i)=O(1)$.  Therefore the instability is  due to the model.

 The paper is organized as follows.
In section~\ref{HSO} we describe  the Hele-Shaw  cell and  the constitutive  equations of 
the Oldroyd-B  fluids.   The basic solution is given   in section~\ref{TBF}.  In  section 
~\ref{TPS}  we get the linear perturbation system. The stability analysis  is   performed 
in section ~\ref{MLSA}. { A special  Fourier decomposition} used in section 
~\ref{FD}  allows us to avoid  the unbounded growth of the  partial derivatives of the 
perturbed velocity near the interface between the two fluids. 
In section  \ref{LOT} we get the leading order terms of the extra-stress tensor and the 
amplitude of the  velocity perturbations. In  section ~\ref{GCF} we obtain the  growth rate 
formula. In section \ref{RES}  we give new rezults on the effects of superficial tension 
in the stability of the displacement process. 
Some dispersion  curves are plotted,  in good agreement with  previous  numerical results. 
We conclude in section    ~\ref{CONC}.
{ The calculations and complex formulas required for exposure are detailed in  
Appendix 1-3.}

  
\vspace{1cm}

\section{ The   Oldroyd-B  fluids}\label{HSO}

{
We consider two  incompressible { immisicble} Oldroyd-B fluids in 
a Hele-Shaw cell parallel with the $xOy$  plane. The cell plates are separated 
by a gap  of thickness $b$.
The cell  length  is  denoted by $l$ and we { use}  the small parameter 
\begin{equation}\label{EPS}
\epsilon  =    b/l <<1.                                      
\end{equation}
}
The fluid 1 is displacing the fluid 2 in the positive direction of the 
 $Ox$  axis. { Across  the sharp interface bewteen the fluids,   
the jump of the  averaged   normal stress should equal  the  surface 
tension  times  the curvature of  averaged  interface and the normal velocity 
should be  continuous } (the    "depth-averaged"    Laplace's  law). The 
no-slip conditions are imposed   on the cell plates.


The velocities, the pressures, the { viscosities} and the extra -
stress tensors for both fluids   $i=1,2$   are denoted by   
$$ { { \underline {\bf u}^i}} = 
({ \underline u}^i,{ \underline v}^i,{ \underline w}^i),  \,\,
{ \underline p}^i, \,\, \mu_i, \,\,  { \underline \tau}^i .                 $$
{
The stress ${\underline \Sigma} ^i$ and the strain-rate ${\bf{\underline S}^i}$
are given  by
} 
$${ { \underline \Sigma} ^i }=
{ \underline p}^i {\bf I} -  { \underline \tau}^i,                           $$
$$   {\bf { \underline S}^i} =
({\bf { \underline V}^i} + {\bf { \underline V}^{iT} })/2,                   $$                                               
$$ { {\bf { \underline V}^i} = \nabla { \underline {\bf u}^i}},              $$
{ where ${\bf I}$ is the unit tensor and $\nabla$   is the gradient 
operator.}

We have the following  flow equations, divergence-free condition  
and  constitutive   relations, with  $c_i>  d_i \geq  0, \,\,\, i=1,2$: 
$$
- {\bf \nabla} { \underline p^i} + {\bf \nabla} \cdot  
{ \underline \tau }^i =0,                                                      $$
\begin{equation} \label{A1}
\quad { \underline u}^i_{x} + { \underline v}^i_y + { \underline w}^i_z =0, 
\end{equation}
\begin{equation}\label{A3}
{\bf { \underline \tau}}^1 + c_1 {\bf { \underline \tau}}^{1 \nabla} =  
2 \mu_1 ( {\bf { \underline S}}^1 + d_1 {\bf { \underline S}}^{1 \nabla} ) ;                                                                            
\end{equation}
\begin{equation}\label{A2}
{\bf { \underline \tau}}^2 + c_2 {\bf { \underline \tau}}^{2  \nabla} =  
2 \mu_2 ( {\bf { \underline S}}^2 + d_2 {\bf { \underline S}}^{2 \nabla} ) .
\end{equation} 
Here $(c_1, d_1), \,\,  (c_2, d_2) $ are the relaxation and the retardation  
(time)  constants of the fluids.
The lower indexes   $_x$, $_y$, $_z$ are denoting  the partial  derivatives;  
${\bf { \underline \tau}}^{ i \nabla},  \quad {\bf { \underline S}}^{ i \nabla} $ 
are the upper convected derivatives.  We consider a steady flow,    then 
$$
{\bf { \underline \tau}}^{ i \nabla}=  
{\bf { \underline u}} ^i\cdot { \nabla}  { \underline \tau}^i - 
({\bf { \underline V}}^i  { \underline \tau}^i + 
 { \underline \tau}^i {\bf{ \underline  V}}^{iT}),\quad                       $$
\begin{equation}\label{A8} 
 {\bf { \underline S}}^{ i \nabla}= 
 {\bf { \underline u}}^i \cdot {\bf\nabla} {\bf { \underline S}}^i - 
({\bf { \underline V}}^i {\bf{ \underline  S}}^i + 
{\bf { \underline S}}^i{\bf { \underline V}}^{iT}).
\end{equation} 

\newpage

\section{The  basic flow }\label{TBF}  

We study the linear stability of the following {\it basic flow},  
denoted by the   super  index $^{0i}$, $i=1,2$:
$$
{\bf \nabla} p^{0i} = ( p^{0i}_{x}(x), 0 , 0), \quad {\bf u}^{0i} = 
( u^{0i}(z), 0, 0 ),                                                  $$
$$
{\bf V}^{0i}_{13}=u^{0i}_z, \quad  {\bf V}^{0i}_{mn}=0, (mn)\neq(13), $$
\begin{equation}\label{A9}
{\bf  S}^{0i} = ({\bf V}^{0i} + {\bf  V}^{0iT} )/2 .            
\end{equation}
{
The basic extra-stress  tensor  $\tau^{0i}$ (depending only on $z$) is 
given by the equations \eqref{A11} - \eqref{A11AC}
in {\bf Appendix 1}  and it follows} (see also \cite{WI}):
\begin{equation} \label{A12A}
\tau^{0i}_{11} = 2 \mu_i (c_i-d_i) (u^{0i}_z)^2;                                                                 
\end{equation} 
\begin{equation}\label{A12}
\tau^{0i}_{13} = \mu_i u^{0i}_z;  \quad
 \tau^{0i}_{mn}=0, \,\, (mn)\neq (11), (13).                       
\end{equation}
{ Therefore we obtain} the following basic flow equations 
$$  p^{0i}_{x} =    \tau^{0i}_{11,{x}}  +  
	\tau^{0i}_{12,y} + \tau^{0i}_{13,z} = \tau^{0i}_{13,z},      $$
\begin{equation} \label{A13}
  p^{0i}_y  =   p^{0i}_z  = 0.
\end{equation}
{ From } the  relation  $\eqref{A13}_1$ { we get }
\begin{equation}\label{A14}
p^{0i}_{x}  =  \tau^{0i}_{13, z} = G_i = \mu_i u^{0i}_{zz} 
\end{equation}
where $G_i$  are two  negative constants.  { We}  suppose   
$u^{0i}= 0$  for $z=0, z=b$,  then we obtain 
\begin{equation}\label{A15}
 u^{0i} = (G_i / 2 \mu_i )z(z-b), \quad i=1,2 .
\end{equation}

The  normal velocity must be continuous across the interface. Our basic 
velocity has only the normal component, {then $u^{01}=u^{02}$ 
and we have the important relation}
\begin{equation}\label{COMPAT}
\frac{G_1}{\mu_1} =   \frac{G_2}{\mu_2} .                                        
\end{equation}
{\it The basic flow velocity $u^0$ } is given by 
\begin{equation}\label{BASIC-U-0}
 u^{0}:= u^{0i} =  (G_i / 2 \mu_i )z(z-b), \quad i=1,2.                                      
\end{equation}
We define the  average  operator 
\begin{equation}\label{AVERAGE}
 \quad <h > = (1/b) \int_0^b h(z)dz 
\end{equation}
and we   introduce the characteristic   velocity $U$:
\begin{equation}\label{COMPAT-0}
 U= <u^{0}> = -\frac{b^2}{12 \mu_i} G_i , \quad G_i = 
-\frac{12 U \mu_i}{b^2}.
\end{equation}
The  relation \eqref{COMPAT-0} is similar with the Darcy's  law  for a  
porous  medium { with permeability $-b^2/12$. }    

We consider  the  following basic  interface between the  displacing 
fluids  
 \begin{equation}\label{COMPAT-00}
x=Ut,
\end{equation}
where $t$ is {\it time} and we introduce the moving reference 
\begin{equation}\label{COMPAT-1A}
{\overline x}=x-Ut.
\end{equation}
Then  the  basic  (material) steady interface   becomes 
\begin{equation}\label{COMPAT-2}
{\overline x} = {\overline x}_0:= 0.                                                    
\end{equation}
 In the following we still use the notation $x$ instead of ${\overline x}$.

It is noteworthy that in  ~\cite{WI}  is considered the planar interface 
$x=Ut$, by using  the  relation  \eqref{COMPAT-0}. 
In \cite {mora2012},  after the formula (10),   it is   specified that 
``since $b$  is much smaller than any lateral lengthscale in a Hele-Shaw 
device, then any $z$ dependence of the basic interface  is not relevant''. 

The flow is due to the pressure gradients. { The pressures contain 
two unknown constants, used latter in order to obtain  the  Laplace's law for 
perturbations.}

\section{The perturbations system }\label{TPS}  

{ The}  linear perturbations of the  basic solution \eqref{A9} are
 denoted  by   
$$ u^i,v^i, w^i, p^i, {\bf V}^i,   {\bf S}^i,  \tau^i, \quad  i=1,2.    $$    
We   assume $ u^i=v^i=w^i=0$ for $z=0,  z=b$.  
The perturbation of the basic interface \eqref{COMPAT-2}  is 
denoted  by $\psi(y,z, t)$ and we have 
\begin{equation}\label{COMPAT-3A}
\psi_t = u^1|_{x_0}= u^2|_{x_0} .
\end{equation} 
The basic velocities verify the divergence-free relation. In  the  frame of 
linear perturbations we obtain   $u^i_x+v^i_y+w^i_z=0$.  
We use the average operator \eqref{AVERAGE} { and get}
$$ <u^i_x+v^i_y >=0.                                                   $$  
This condition is  verified if   $u^i_x+v^i_y=0$, from which we get  $w^i_z=0$.
{   In this   paper we  consider }
\begin{equation}\label{COMPAT-3B}
u^i_x+v^i_y=0,  \quad w^i=0. 
\end{equation}

{ The  following  relations  \eqref{P1} -   \eqref{P4} are concerning 
both fluids 1 and 2}, then   we omit the super index  $^i$. We introduce the small 
perturbations in \eqref{A3}-\eqref{A8} and obtain 
$$
{\bf \tau}^0 + {\bf \tau} + \theta ({\bf \tau}^0 + {\bf \tau})^{{\bf \nabla}} =  $$
\begin{equation}\label{P1}
2 \mu [{\bf S}^0 + {\bf S} + \eta ({\bf S}^0 +{\bf S})^{\bf \nabla}],
\end{equation}
$$
({\bf \tau}^0 + {\bf \tau})^{{\bf \nabla}} = 
u^0 {\bf \tau}_x-[{\bf V}^0{\bf \tau}^0+{\bf \tau}^0 {\bf V}^{0T}]-             $$
\begin{equation}\label{P2}
[{\bf V}^0 {\bf\tau} + 
{\bf V} {\bf \tau}^0 + {\bf \tau}^0 {\bf V}^T + {\bf \tau} {\bf V}^{0T}],
\end{equation}
$$
({\bf S}^0 + {\bf S})^{{\bf \nabla}} =  u^0 {\bf S}_x - 
[{\bf V}^0 {\bf S}^0 + {\bf S}^0 {\bf V}^{0T}] -                                 $$
\begin{equation}\label{P3}
[{\bf V}^0 {\bf S} + {\bf V} {\bf S} ^0 + 
{\bf S}^0 {\bf V}^T + {\bf S} {\bf V}^{0T}]
\end{equation}
where $\mu, \theta, \eta$ verify the relations
 \begin{equation}\label{SALT-MIU}
\mu = \mu_1 \mbox{ for }    x < x_0;  \quad 
\mu = \mu_2 \mbox{ for }    x > x_0;                                                                            
\end{equation}
\begin{equation} \label{SALT-THETA-ETA}
\theta = \left \{   \begin{array} {c} 
 c_1,  x <x_0 \\
 c_2,  x >x_0   
\end{array} \right. ;                 \quad 
\eta = \left \{\begin{array} {c} 
d_1,  x<x_0 \\
d_2,  x>x_0 \\ 
\end{array} \right.  .
\end{equation} 
In the frame  of the linear stability (by neglecting the second order terms in 
perturbations)   it follows 
\begin{equation}\label{P4}
{\bf \tau} + \theta ( u^0 {\bf \tau}_x- {\bf E})=
\mu\{ 2{\bf S} + \eta(  u^0 {2\bf S}_x - {\bf F})\},
\end{equation}  
{
where the  tensors  ${\bf E}, {\bf F}, {\bf S}$ are given  in {\bf Appendix 2} - 
see the relations \eqref{P6}-\eqref{P7}.
}

The perturbed  normal   stresses  in both fluids are  given by  (see  ~\cite{RE},  
~\cite{BU}) 
$$ T^i_{11} = p^{0i} + p^i - (\tau^{0i}_{11}+ \tau^i_{11} ).                 $$
We search   for the  limit values  
$ T^-_{11}= lim_{ \,\, x<x_0, \,\,  x\rightarrow x_0 \,\,}T^1_{11} \quad $ and 
$ \quad  T^+_{11}= lim_{ \,\, x>x_0, \,\, x\rightarrow x_0 \,\,}T^2_{11}$. 
 
{ The} basic pressure is not depending on $z$, { then} we 
have  the following  first order  expansion  near the  basic interface $x_0$:
$$
p^{0i}(x_0+ \psi) =  p^{0i}(x_0+ <\psi>)  =                                  $$
\begin{equation}\label{COMPAT-5}
   p^{0i}(x_0) + p^{0i}_x(x_0) <\psi> =   
p^{0i}(x_0) + G_i <\psi>.                        
\end{equation}                                                                                                                      
Recall \eqref{COMPAT-3A}, then $\psi$    is continuous across the interface 
 $x=x_0$   and we get
$$
T^-_{11}  = p^{01}(x_0)+ G_1 <\psi> + p^1 -  (\tau^{01}_{11}+ \tau^1_{11} ), $$
 \begin{equation}\label{COMPAT-6} 
T^+_{11} = p^{02}(x_0)+ G_2 <\psi> + p^2 -   (\tau^{02}_{11}+ \tau^2_{11} ).                                        
\end{equation}   
The  depth-averaged  dynamic  Laplace's law (near the basic interface $x=x_0$) 
is  
$$ < T^+_{11}- T^-_{11} >=  \gamma \times <interface \,\, curvature >        $$  
{ where $\gamma$ is the surface tension. From }  \eqref{COMPAT-5}, 
\eqref{COMPAT-6} it follows                           
$$   p^{02}(x_0) + G_2 <\psi> +  
{ < p^2 } - \tau^{02}_{11}+ \tau^2_{11} > -                       $$
$$\{ p^{01}(x_0) + G_1 <\psi> +  
{ < p^1 } - \tau^{01}_{11}+ \tau^1_{11} > \}=                     $$
\begin{equation}\label{COMPAT-6A}
\gamma(x_0) < x_{0yy} + x_{0zz} +  \psi_{yy} + \psi_{zz} >,                                             
\end{equation}
where { the curvature  of $\psi$  is approximated by  
$(\psi_{yy}+\psi_{zz})$.}  
The basic normal stress verify the Laplace's law on the basic interface  $x=x_0$, 
so we should have the relationship
$$  (p^{02} - <\tau^{02}_{11}>)|_{x_0} -  (p^{01} - <\tau^{01}_{11}>)|_{x_0}= $$
\begin{equation}\label{COMPAT-6A0}
\gamma(x_0) < x_{0yy} + x_{0zz} >=0.        
\end{equation}
Only the basic  pressures gradients  are given, then the basic pressures contain
two additive constants.  As in \cite{WI} (where the displacing fluid is air), for 
appropriate values of these constants we get the above relation - { see
also the last two lines of section   \eqref{TPS}}.  
Therefore the relations  { \eqref{COMPAT-6A}-\eqref{COMPAT-6A0}  are  }
giving us  the  Laplace's law for   perturbations:  
$$
 (G_2-G_1) <\psi> + { <p^2 - \tau^2_{11}> - <p^1 - \tau^1_{11}> }  $$
\begin{equation}\label{COMPAT-7}
 = \gamma(x_0) < \psi_{yy} + \psi_{zz} >.
\end{equation}

\section{Linear stability analysis }\label{MLSA}

\subsection{ Fourier decomposition}\label{FD} 

We consider  the following  expansions of the velocity  perturbations, 
with  { $k \geq 0, \quad \alpha >0 $ }:
$$
u^1 =  f(z) E1 \cos(ky),     \quad                               $$
\begin{equation}\label{SA1}
v^1 = - f(z)  E1 \sin(ky),    \quad  x  < x_0;                                                            
\end{equation}   
$$
u^2 =  f(z) E2 \cos(ky),      \quad                               $$
\begin{equation}\label{SA2}
v^2 =   f(z)   E2 \sin(ky),    \quad x >x _0;                                                                  
\end{equation}  
\begin{equation}\label{SA2A}
 f(z) = \beta u^0(z), \quad \beta = O(\epsilon^2);                                                                  
\end{equation}  
$$
E1= \exp( - k \alpha + kx + \sigma t ),  \,\, x \leq x_0;          $$
\begin{equation}\label{SA3}
E2= \exp( - k \alpha - kx + \sigma t ),  \,\, x \geq x_0.
\end{equation}  
The dimension of $\alpha $ is { \it length}.  On $x_0$  we have  
$$
     u^1=u^2, \quad v^{1-}\neq v^{2+}, \quad                        $$
$$ v^1_y = -kf E1|_{x_0} \cos(ky), \quad                            $$
\begin{equation}\label{SA4}
v^2_y =  kf E2 |_{x_0}\cos(ky), 
\end{equation}
where $^-, ^+$ are the lateral limits  values.  
The perturbations  decay to zero far from { the}  interface  $x=x_0$ and 
$u^i_x+v^i_y=0$  in both fluids. 
If $\alpha \neq 0$, then $u_x, u_y, v_x, v_y$   { near $x_0$}
are bounded in  terms of $k$.
{ We justify the form of the amplitude $f$ in  section 5.2}.


The { perturbation of the basic} interface was denoted by 
$\psi(x,y,z,t)$ {- see} \eqref{COMPAT-3A}. The Fourier decomposition  
\eqref{SA1}-\eqref{SA3} gives us $\psi=u/\sigma$ and from \eqref{COMPAT-7}  
it follows    
\begin{equation}\label{COMPAT-7A1}
\sigma = \frac{ \gamma <u_{yy} +  u_{zz}> +(G_1-G_2) <u> }
 { <p^2  - \tau^2_{11}> - <p^1  - \tau^1_{11}> }
\end{equation}    
where $u=u^1(x=0)=u^2(x=0)$.
{ We have to obtain } $<p^i  - \tau^i_{11}> $ as a functions 
 of basic and perturbed   velocities. 
{ In this paper we use} the following dimensionless quantities, 
denoted by $'$ : 
$$ x'=x/l, \quad y' = y/l, \quad z'=z/b,                                  $$
$$ u^{'i}=u^i/ U, \quad v^{'i}   =v^i/ U,                                 $$
$$  p^{'i}=p (l/\mu_i U), \quad  \gamma'= \gamma /(\mu_1 U),   \quad  
\mu_0= \mu_2/\mu_1 ,                                                      $$
$$ \alpha' = \alpha/ l, \quad  k'=kl,  \quad  \sigma' = \sigma / (U/l), 
\quad t' = t(U/l).                                                    $$
$$  \{ (\tau^i _{11})' , (\tau^i_{ 12})',   (\tau^i_{ 22})' \} =  
\{ \tau^{i}_{11}  ,\tau^{i}_{ 12},   \tau^{i}_{ 22}  \} (l/\mu_i U),       $$         
$$  \{ (\tau^i_{13})' ,(\tau^i_{ 23})',   (\tau^i_{ 33})' \} = 
\{ \tau^{i}_{13}  ,\tau^{i}_{ 23},   \tau^{i}_{ 33}  \} (b/\mu_i U),       $$
\begin{equation}\label{COMPAT-9} 
C_i = c_i U/b,   \quad D_i= d_i U/b.                                                                    
\end{equation}
{  Here $C_i,  D_i$ are the Weissenberg numbers.  
We consider that  $C_i,D_i, \exp(\sigma' t')$ are of order $O(1)$}. 

\begin{omitext}
{ We have the    following lateral limit values}:  
$$
\psi(x)^-_x(x_0)=                                                       $$ 
\begin{equation}\label{INTER001A}
k \frac{f(z)}{\sigma} \exp(-k \alpha + \sigma t )\cos(ky), 
\end{equation}
$$
\psi(x)^+_x(x_0)=                                                       $$
\begin{equation}\label{INTER001}
 -k \frac{f(z)}{\sigma} \exp(-k \alpha + \sigma t )\cos(ky). 
\end{equation}
\end{omitext}   



\subsection{Leading order  terms   for $\tau, p$}\label{LOT}

In this subsection we use only the  dimensionless   quantities,  but  we omit   
the $'$. We obtain approximate formulas of $\tau_{ij}, p_z, p_x, p_y$  in terms 
of  $u^0, u, v$. 

{  The leading order expressions for all components of  $\tau$ are 
given in {\bf Appendix 3}.  For both fluids ($i=1,2$)  we have }
\begin{equation}\label{PRIMELE-3-AP}
 \tau^i_{33}=0,  \, \tau_{31}^i=u^i_z   , \,  \tau^i_{32}=v^i_z,
\end{equation}
\begin{equation}\label{TAU-12-AP} 
  \tau^i_{12,y}  =  
  (u^i_y+v^i_x)_y  + 2  (C_i -D_i) u^0_{z} v^i_{zy}/ \epsilon,   
\end{equation} 
\begin{equation}\label{TAU-11-AP}
 \tau^i_{ {11,x}} =   2u^i_{xx}  + 4  (C_i-D_i) u^0_zu^i_{zx}/ \epsilon   . 
\end{equation}
 		

We get  $p^i_z, p^i_x$  by using  { the flow equations} and 
the divergence-free condition:  
\begin{equation}\label{PZ}
p^i_z = \tau^i_{31,x}+\tau^i_{32,y} =  u^i_{zx}+ v^i_{zy}  = 0;
\end{equation}
$$
p^i_{x} = \tau^i_{11,x}+\tau^1_{12,y}+\tau^i_{13,z} / \epsilon^2 =       $$
$$ 2  u^i_{xx} + 4 (C_i-D_i)u^{0}_{z}u^i_{zx} / \epsilon  +              $$ 
\begin{equation}\label{PX-A}
(u^i_{y}+v^i_{x})_{y} +  2 (C_i-D_i)u^{0}_{z} v^i_{zy} / \epsilon + 
 u^i_{zz} /  \epsilon^2 .                                                           
\end{equation}
 We also have  $ u^i_{zx} + v^i_{zy} =0, \quad u^i_{xx}+ v^i_{xy}=0, \quad
{ u^i_{xx} + u^i_{yy} =0} $ and   from  \eqref{PX-A} it follows  
\begin{equation}\label{PX}
p^i_{x} =  2(C_i-D_i)u^{0}_{z} u^i_{zx} / \epsilon  +  u^i_{zz} / \epsilon^2.
\end{equation}
On the same way we get
\begin{equation}\label{PY}
p^i_{y} =  2(C_i-D_i)u^{0}_{z} v^i_{zx} / \epsilon  +  v^i_{zz} / \epsilon^2.
\end{equation} 
As $u^i_{zxx}+ v^i_{zxy}=0, \,\, u^i_{zzx}+v^i_{zzy}=0$,   from \eqref{PX} - 
\eqref{PY}  we obtain
\begin{equation}\label{PYY}
p^i_{xx}   + p^i_{yy} =0.
\end{equation}

 The form of the amplitude $f$ is justified as follows. { 
From  \eqref{PZ}, \eqref{PX} we should have }  
\begin{equation}\label{PXZ=0}  
 p^1_{xz}=  2 ( C_1-D_1) (u^0_z u^1_{zx})_z/ \epsilon  
 + u^1_{zzz}/ \epsilon^2 =0.           
\end{equation}
{We prove that $p^1_{xz} =O(\epsilon)$. For this, we recall  
$ \beta = O( \epsilon^2)$ and we use the inequality }        
$$max_k \{u^1_x\}  \leq  max_k \{ k \exp(-k \alpha + kx) \} \leq   $$
$$ max_k \{ k \exp(-k \alpha) \} = { 1/(\alpha e),}                $$ 
{ which holds for $x<0$ and  $k>0$.} Then
$$ { (u^0_z u^1_{zx})_z/ \epsilon } =\beta O(1)/\epsilon = 
O(\epsilon) ,   \quad      
 ( C_1-D_1) (u^0_z u^1_{zx})_z / \epsilon  =O(\epsilon).            $$ 
The decomposition \eqref{SA1}-\eqref{SA3} is giving us $u^1_{zzz}=0$. 
Then
$f(z)=\beta u^0(z)$ verifies  \eqref{PXZ=0}  with  the precision order  
$ O(\epsilon) $, if $\alpha$ is large enough.

\section{The growth rate formula}\label{GCF}

In this section we use   both dimensional and dimensionless quantities (the last 
are denoted by  $^{'}$) and obtain the growth-rate formula.  
{ The   flow  equations ad the decomposition \eqref{SA1}-\eqref{SA3}
give us}
$$
p^1 - \tau^1_{11} = \frac{1}{k}(\tau^1_{12,y} + \tau^1_{13,z} ), \,  x<x_0; $$      
\begin{equation}\label{P-TAU}
p^2 - \tau^2_{11} = \frac{-1}{k} ( \tau^2_{12,y} + \tau^2_{13,z} ), \,  x>x_0.                  
\end{equation}
{
The dimensional forms of the equations \eqref{PRIMELE-5}, \eqref{TAU-12-F} in  
{\bf Appendix 3} are
 $$
\tau^i_{12,y} = \mu_i[(u^i_y + v^i_x)_y +  2(c_i-d_i) u^{0}_z v^i_{zy}],   $$
\begin{equation}\label{TAU-12-Y}
\tau^i_{13}= \mu_i u^i_z, \,\, \tau^i_{23}= \mu_i v^i_z, \,\,\tau^i_{33}= 0. 
\end{equation}
}
The relations  \eqref{SA1}, \eqref{SA2},  \eqref{SA4} and \eqref{P-TAU} -  
\eqref{TAU-12-Y}  give us
$$  (u^1_y+v^1_x)_y = -2 k^2 f(z) E1 \cos (ky);   \quad                 $$    
$$  (u^2_y+v^2_x)_y = -2 k^2 f(z) E2 \cos (ky);                         $$
$$ p^1 - \tau^1_{11} = \frac{1}{k}    \mu_1 [(u^1_y+v^1_x)_y + 
2 (c_1-d_1) u^{0}_zv^1_{zy} + u^1_{zz}  ]=                              $$
\begin{equation}\label{P-T-1}
   \frac{1}{k}\mu_1[ -2k^2 f { -}  
	2k(c_1-d_1)u^{0} f_z + f_{zz} ] E1 \cos(ky);                          
\end{equation}
$$ p^2 - \tau^2_{11} = \frac{- 1}{k}  \mu_2 [(u^2_y+v^2_x)_y + 
2 (c_2-d_2) u^{0}_zv^2_{zy} + u^2_{zz}  ]=                               $$
\begin{equation}\label{P-T-2}
  \frac{-1}{k}\mu_2[-2k^2 f +  2 k(c_2-d_2) u^{0} f_z  + f_{zz} ] E2 \cos(ky).   
\end{equation}
{ We use  \eqref{COMPAT-7}, \eqref{P-T-1}, \eqref{P-T-2} and we 
 get } 
$$ (G_2-G_1) \frac{<u^0>}{\sigma} -    
\frac{\gamma}{\sigma} <-k^2 u^0 + u^0_{zz}> +                            $$
$$ -\frac{1}{k}  <-2k^2 u^0 M + 2k N(u^0_z)^2 +                               
   u^0_{zz} M > =0,                                                      $$
$$ M= (\mu_2+\mu_1),     \,\, M'= (\mu_0+1),                             $$
$$  N =    \mu_2(c_2-d_2) -\mu_1(c_1 - d_1),                             $$      
\begin{equation}\label{COMPAT-08}
  N' =    \mu_0(C_2-D_2) -(C_1 - D_1).          
\end{equation}
{ From \eqref{A15} - \eqref{COMPAT-0} we obtain  the following 
averages }
$$  <u^0>=U,                                                             $$
$$  <(u^0_z)^2>= 12 U^2/b^2 ,                                            $$
$$  <u^0_{zz}>= - 12 U/b^2,                                              $$
{ then $\eqref{COMPAT-08}_1$ leads us to}:
\begin{equation}\label{COMPAT-8}
 \sigma =   \frac{ U k (\mu_2-\mu_1) - \gamma (k^3 b^2 /12 +  k)} 
{ M k^2 b^2 /6 - 2k  N U + M }.  
\end{equation}
Let $Ca= (\mu_2-\mu_1) U/ \gamma $  be the capillary number, then the relation 
\eqref{COMPAT-8} becomes}
\begin{equation}\label{COMPAT-8A}
 \sigma =   \frac{ U k (\mu_2-\mu_1)(1- Ca^{-1}) - \gamma (k^3 b^2 /12)} 
{ M k^2 b^2 /6 - 2k  N U + M }.  
\end{equation}

The dimensional  Saffman-Taylor formula is
\begin{equation}\label{S-T}
 \sigma_{ST}                    =                                                                
 \frac{Uk(\mu_2- \mu_1) - \gamma ( k^3 b^2/12)} { M }.
\end{equation}

When $Ca >> 1$  we have  the same numerators in   
\eqref{COMPAT-8A} and \eqref{S-T}; only the  denominator of \eqref{COMPAT-8A}   
contains the two new terms { $(M k^2b^2/6 - 2k N U)$ instead of $M$}.

From \eqref{COMPAT-9}, \eqref{COMPAT-08}, \eqref{COMPAT-8} and \eqref{S-T}   
we get  the   dimensionless   expressions
\begin{equation}\label{COMPAT-10}
 \sigma' =   \frac{  k' (\mu_0-1) - \gamma' (k'^3 \epsilon^2 /12 +  k')} 
{ k'^2 M' \epsilon^2 /6  - 2k' N'\epsilon  + M'},    
\end{equation}
\begin{equation}\label{COMPAT-11}
\sigma'_{ST}  =  \frac{ k{'}(\mu_0- 1) - 
\gamma' ( k'^3 \epsilon^2/12)} {M'}.
\end{equation}

\vspace{0.25cm}						
						
{\it  Remark 1.}  We have  						
$$
<\psi>=   { <u>/ \sigma }=                                 $$ 
\begin{equation}\label{REMARK-1}
\frac{1}{\sigma} < f(z) > \exp(-k \alpha ^+_- kx + \sigma t) \cos (ky).                                 
\end{equation}
Then, near the  basic interface, $p^{0i}$ given by   \eqref{COMPAT-5} is 
depending  on $y$.  But our basic pressures   must depend only on $x$ - 
see  \eqref{A13}.  We can partially overcome   this inconsistency by using 
the parameter $\alpha$.  For this, we estimate the partial derivative  of 
the  perturbed interface { with respect to $y$}.  As we mentioned 
at the end of section 5.1, we  consider $\exp(\sigma' t' ) = O(1)$.  Then 
we have  
$$
  <\psi>_y  \leq   \frac{1}{\sigma}  f(z) F(k),                          $$
$$  F(k):=   k \exp( - k \alpha)\quad  { \leq } \quad 
\frac{1}{\alpha e}.                                                      $$                         
A large enough  $\alpha$ is giving  us  an arbitrary  small $p^{0i}_y$. 
If $\alpha =0$ in the decomposition  \eqref{SA1}-\eqref{SA3}, then for
$x \rightarrow 0$   and $k \rightarrow \infty$ we get  
$ <\psi>_y, |u_x|, \, |u_y|, \, |v_x|, \, |v_y|  \rightarrow \infty$.
\hfill  $\square$

\vspace{0.5cm}						
						
{\it  Remark 2.}  Our model can describe the displacement of an Oldroyd-B fluid 
by air. For this, we consider  \eqref{COMPAT} in the form $G_1= \mu_1 G_2/ \mu_2$.
{ As the displacing fluid is air, then   $\mu_1 \approx 0$  and  from  
\eqref{COMPAT},  \eqref{P-TAU},   \eqref{TAU-12-Y}  we get} 
$$ G_1 =0,   \quad p^1 - \tau^1_{11}=0.                                        $$
In this case  the Laplace's law \eqref{COMPAT-7}  becomes  	
\begin{equation}\label{COMPAT-12}
 G_2 <\psi> +   <p^2 -  \tau^2_{11}>  =  \gamma(x_0) <\psi_{yy} + \psi_{zz}>.
\end{equation}   
This form of the Laplace's law (by  neglecting  the meniscus  curvature  $\psi_{zz}$)
was used in ~\cite{WI}, based on  the  additional hypothesis
$$ p^{02} = G_2(x-Ut), \quad x >x_0,                                            $$
which in fact it's not   necessary.  
If $\mu_1=0, c_1=d_1=0$ then the formula \eqref{COMPAT-8} becomes 
\begin{equation}\label{COMPAT-8-BB}
 \sigma =   \frac{ U k \mu_2 - \gamma (k^3 b^2 /12 +  k)} 
{ \mu_2 [ k^2 b^2 /6 +1] - 2k   \mu_2(c_2-d_2) U}.
\end{equation}
In $\eqref{COMPAT-9}_3$ we put  
$ \gamma'=\gamma /(\mu_2 U)  \mbox{ instead of }  \gamma'=\gamma /(\mu_1 U)$.
{ Moreover, $\mu_0=1$. Then  we get the  dimensionless  growth rate}
\begin{equation}\label{COMPAT-10A}
 \sigma'_{AIR} =   \frac{  k'  - \gamma' (k'^3 \epsilon^2 /12 +  k')} 
{ k'^2 \epsilon^2 /6  - 2k'  (C_2-D_2) \epsilon  + 1}.   
\end{equation}
The dimensionless Saffman-Taylor  formula   is 
 \begin{equation}\label{COMPAT-10B}
 \sigma'_{ST-AIR} =   k'  - \gamma' k'^3 \epsilon^2 /12.
 \end{equation} 
\hfill  $\square$


\section{ Discussions  and  results }\label{RES}

 A.  We consider 
$C_i=D_i=0,\,\, or \,\,  C_i=D_i,  \,\, i=1,2 \quad   or  \quad  
(C_1-D_1)=  \mu_0 (C_2-D_2) $. 
Then the growth constant    \eqref{COMPAT-10} becomes
\begin{equation}\label{COMPAT-13}
 \sigma' =   \frac{  k' (\mu_0-1) - \gamma' (k'^3 \epsilon^2 /12 +  k')} 
{ (\mu_0+ 1) [ k'^2 \epsilon^2 /6 +1]  },    
\end{equation}
{ therefore  $\sigma' < \sigma'_{ST}$.}
 Two new  terms  appear in   \eqref{COMPAT-13}, compared with  the formula  
\eqref{COMPAT-11}: 

 i) $  (\mu_0+ 1)k'^2 \epsilon^2/6 $   in the denominator; \hspace{0.5cm}   
ii)  $ - k'  \gamma' $   in  the   numerator.

\noindent 
 The dispersion  curves are  given in Figure 1.  The new terms in the formula 
\eqref{COMPAT-13} appear from  two reasons:

 a) we not neglected   $u_x,v_x, u_y, v_y$ in front of $u_z, v_z$;

 b) we used the {\it total} curvature of the perturbed interface in the 
Laplace law \eqref{COMPAT-7}.

As a consequence,  we obtain the following results:   

A1) { \it Even if the surface tension $\gamma'$ on the interface is zero,
the growth  constant is bounded in terms of the wave  number $k'$}.  The 
equation  \eqref{COMPAT-13} with $\gamma'=0$ gives   us 
$$ 
 \sigma' = R \frac{ k'} {(1 +  k'^2 \epsilon^2/6)} <    
           R \frac{(1+ \delta) \sqrt 3}{ \epsilon \sqrt 2}, \quad 
\forall \delta >0,                                                   $$
\begin{equation}\label{FINAL-T-0}  
 \quad R = \frac{(\mu_0-1)}{(\mu_0+ 1)}.                                 
\end{equation}
Indeed, we have  
$$ \frac{ k'}{ 1+ k^{'2} \epsilon^2/ 6 } < B,   \,\, \forall \,\, 
k' \geq 0                                                           $$
$$ \Leftrightarrow   
1- \frac{2 \epsilon^2}{3} B^2 <0    \Leftrightarrow  
B >  \frac{ \sqrt 3}{ \epsilon \sqrt 2}.                            $$
 \hfill  $\bullet$                                    

A2) If the surface tension $\gamma'$ on the interface is zero, then 
the growth  constant tends to zero for very large wave numbers $k'$.

We cite here some results obtained for displacements of immiscible Newtonian 
fluids with very small (or zero) surface tensions on the intrerface, in 2D 
Hele-Shaw cells.In \cite{VAM} (Introduction)  it is specified that
"One asks whether a non-zero-surface-tension model approximates the 
zero-surface-tension one. The answer is negative in the case of a receding 
fluid (see numerical evidence in  \cite{CENI}, \cite{QIN}).  In the case
of injection the answer is supposed to be affirmative but is still unknown".
In \cite{LIN} is  given a perturbation theorem for strong polynomial 
solutions  to the zero surface tension Hele-Shaw equation driven by injection
or suction, the so called  Polubarinova - Galin equation.  In the  case  of 
suction, by using some additional hypothesis, it is proved { that}
the most {part} of the  fluid will be sucked before the strong 
solution blows up.

The above results A1) and A2) are  in contradiction with the Saffman-Taylor 
formula \eqref{COMPAT-11}, where $\gamma'=0$ is giving an  unbounded  
growth constant in terms of  $k'$.
  \hfill  $\bullet$     

A3) From   \eqref{COMPAT-13} we get
\begin{equation}\label{FINAL003A}
\sigma' < \frac{k' \{(\mu_0-1) - \gamma' \}}  {(\mu_0+1)(1 +  k'^2 \epsilon^2/6)}
\end{equation}
and  it follows
\begin{equation}\label{FINAL003B}
 \gamma' > \mu_0-1 >0   \Rightarrow    \sigma'  \leq 0.
\end{equation}
The  growth constant  is negative or zero when  the surface tension $\gamma'$ 
is large enough,  even if the displacing fluid is  less viscous (that means 
$\mu_2>\mu_1$). This is also in contradiction with the Saffman-Talor criterion 
derived from \eqref{S-T}. 
This is an  important result of our paper: the displacement stability  in a 3D 
Hele-Shaw cell  is decided not only by the { ratio of the viscosities
of the  two fluids}, 
but  also by the surface tension on the interface. When the displacing  fluid 
is less viscous,  the sufficient   condition for the {\it almost  stability} is
\begin{equation}\label{FINAL003C} 
 \gamma' > \mu_0-1. 
\end{equation}
A quite similar result is given by the formula (19)  of  \cite{MI}: the growth-rate can not 
be positive for large enough surface tension. But in our formula  \eqref{FINAL003C}  we have 
also the  viscosities ratio. 

\hfill   $\bullet$

A  different  contradiction of the Saffman  and Taylor stability criterion was 
observed in ~\cite{CL},  ~\cite{CHAN},  ~\cite{FKH}, ~\cite{GU1}, ~\cite{GU2}, 
 ~\cite{KH}. All these papers are  related with the displacement of air (then 
$\mu_2$ is   almost zero)  by a fluid 
with surfactant properties in a  Hele - Shaw cell with  preexisting surfactant layers 
on the plates; it is   pointed out  that a  more viscous displacing  fluid can give 
us an {\it unstable}  air-fluid interface.  The experiments and  the numerical  
results  are in good agreement - but also in a   3D frame. We can consider that our 
result is a complementary one, compared with the  above experiments  with surfactants 
fluids and Hele-Shaw cells. We proved  that for a large enough surface tension, even 
if the displacing fluid  is less viscous, the interface   air-fluid is {\it almost 
stable}.   
\hfill   $\bullet$ 

\vspace{0.5cm}

 B.  Consider now  the case when at least one of the  Weissenberg numbers  is not 
equal to  zero and   $(C_1-D_1) \neq  \mu_0 (C_2-D_2) $.

B1)  Let    $(C_i-D_i) \neq 0, \, i=1,2$.
We { use} the notations  \eqref{COMPAT-08} and  introduce { 
the   new quantity   $\Delta$:} 
\begin{equation}\label{FINAL003D}  
 \Delta = (N')^2  -(M')^2 /6.                                            
\end{equation}
In the formula  \eqref{COMPAT-10}  we must avoid the critical value 
\begin{equation}\label{FINAL003F} 
 (N'_{cr})^2     =  (M')^2 / 6.   
\end{equation}
The denominator of the growth rate \eqref{COMPAT-10} is strictly positive 
in the range $\Delta <0$.  As a consequence,   from \eqref{COMPAT-10}  
we get the following {\it instability criterion}: 
$$
0< \gamma' <(\mu_0-1)   \mbox{   and   } 
   (N') ^2   < (\mu_0+1)^2 /6   \Rightarrow                 $$
\begin{equation}\label{COMPAT-14}
  \sigma'   > 0 \mbox{ for }  
(k')^2  < 12 (\mu_0-1- \gamma')/( \epsilon^2 \gamma' ).
\end{equation}  
Moreover,  when 
$$  (N')   ^2   \leq (\mu_0+1)^2/6,                                       $$  
the  denominator 
of  \eqref{COMPAT-10} is close to zero and we get a  blow-up of the growth  
rate. Then a strong destabilizing effect appears, compared with the case of
Newtonian displacing fluids.     

We have also
$$
(\mu_0-1) <0    \mbox{  and } 
 (N') ^2    < (\mu_0+1)^2 / 6   \Rightarrow     $$
\begin{equation}\label{COMPAT-15}
  \sigma'  \leq  0 .                                                                                             
\end{equation}  
 
B2) If  $C_2-D_2 \neq 0$ and $ (C_1 - D_1) =0 $ or $C_1=D_1=0$  we have a 
Stokes displacing  fluid.  In this case, with $ N'= [\mu_0 (C_2-D_2)] $ 
 we  recover  the above results   \eqref{COMPAT-14} -  \eqref{COMPAT-15}.

  \hfill  $\bullet$

We consider a Stokes fluid (so  $C_1=D_1=0$)  displacing an Oldroyd-B fluid with the  
Weissenberg numbers  $C_2-D_2\neq 0$.   
{ In Figures 2,3   we compare   our dispersion curves  \eqref{COMPAT-10}
and  Saffman-Taylor   formula \eqref{COMPAT-11}, }
in the case  $ \gamma'=0.1, \,\, \epsilon=0.006$, for $\mu_0=2$ and $\mu_0=4$. The maximum 
value of  $\sigma'$  given by  \eqref{COMPAT-10} 
is increasing as function of $(C_2-D_2)$ until  the blow-up appears, 
for a finite value of $(C_2-D_2)$ which we denote by  $(C_2-D_2)_{cr}$. The formula 
\eqref{FINAL003F}  gives us
 \begin{equation}\label{COMPAT-B1}
(C_2-D_2)_{cr} = \frac{\mu_0+1}{\mu_0 \sqrt 6 }> \frac{1}{\sqrt 6 }, \quad 
\mu_0 = \frac{\mu_2}{\mu_1}.
\end{equation}
If the ratio $\mu_0$ is increasing, then the critical numbers for which the blow-up of the 
growth rate appears is decreasing. For $\mu_0 \rightarrow \infty$ we get $(C_2-D_2)_{cr} 
\rightarrow 1/\sqrt 6$. 
 This is natural:  if   the viscosity of the displacing  fluid  is decreasing to zero
   then the blow-up of $\sigma'$  appears "earlier", for smaller values of 
	$(C_2-D_2)_{cr} $.


{ In Figures  4, 5  are plotted } 
 the growth rates  \eqref{COMPAT-10A}   when air 
(then $C_1=D_1=\mu_1=0$) is displacing an Oldroyd-B fluid.
In Figure 4 we compare  \eqref{COMPAT-10A}  and \eqref{COMPAT-10B} for   $ \gamma'=0.1, 
\quad \epsilon=0.006$. The maximum value  of  
$\sigma'$ is increasing in terms of $(C_2-D_2)$ until we get the blow-up of the growth   
rate for the critical value    
 \begin{equation}\label{COMPAT-B3}
(C_2-D_2)_{cr} = \frac{1}{\sqrt 6} .
\end{equation}
In Figure 5 { are plotted} the  growth rates \eqref{COMPAT-10A} when $C_2= 0.375$ 
and  $r= D_2/C_2$,  for $r$= 1, 0.7, 0.5, 0.3,  0.1, 0.
 The   dispersion curves given in Figures 4,5  are quite similar with the  numerical 
results  given in Figures 1, 3 of  \cite{WI}.

\newpage

\begin{figure}[!h]
\centering
\vskip 0.2truecm
\includegraphics[scale=0.4]{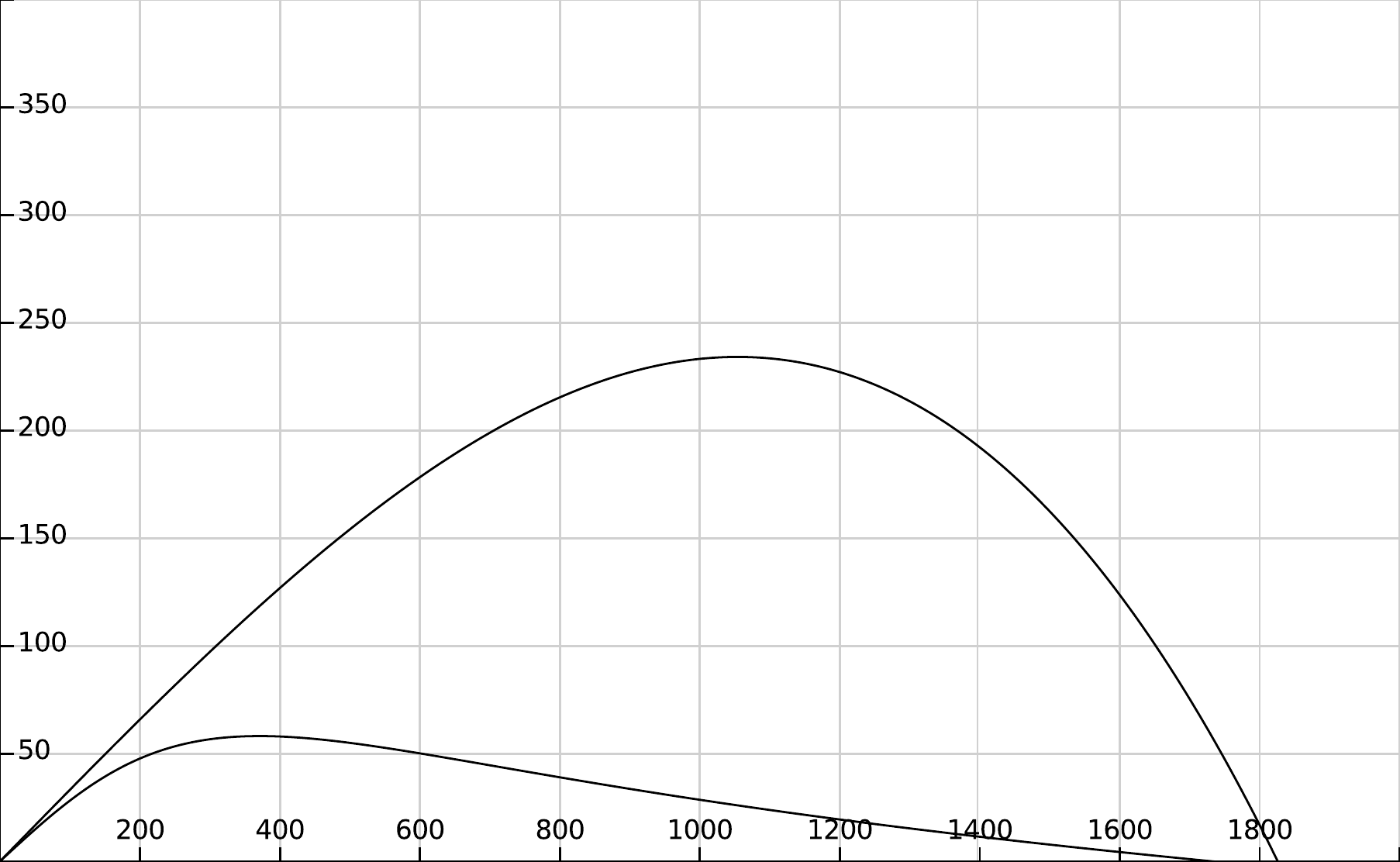}
\caption{Dispersion curves \eqref{COMPAT-11}(upper)   and  \eqref{COMPAT-13}  
for  $\gamma '  =0.1, \epsilon=0.006, \mu_0=2$}
\label{F1}
\end{figure}

\vspace{0.5cm}

\begin{figure}[!h]
\centering
\vskip 0.2truecm
\includegraphics[scale=0.4]{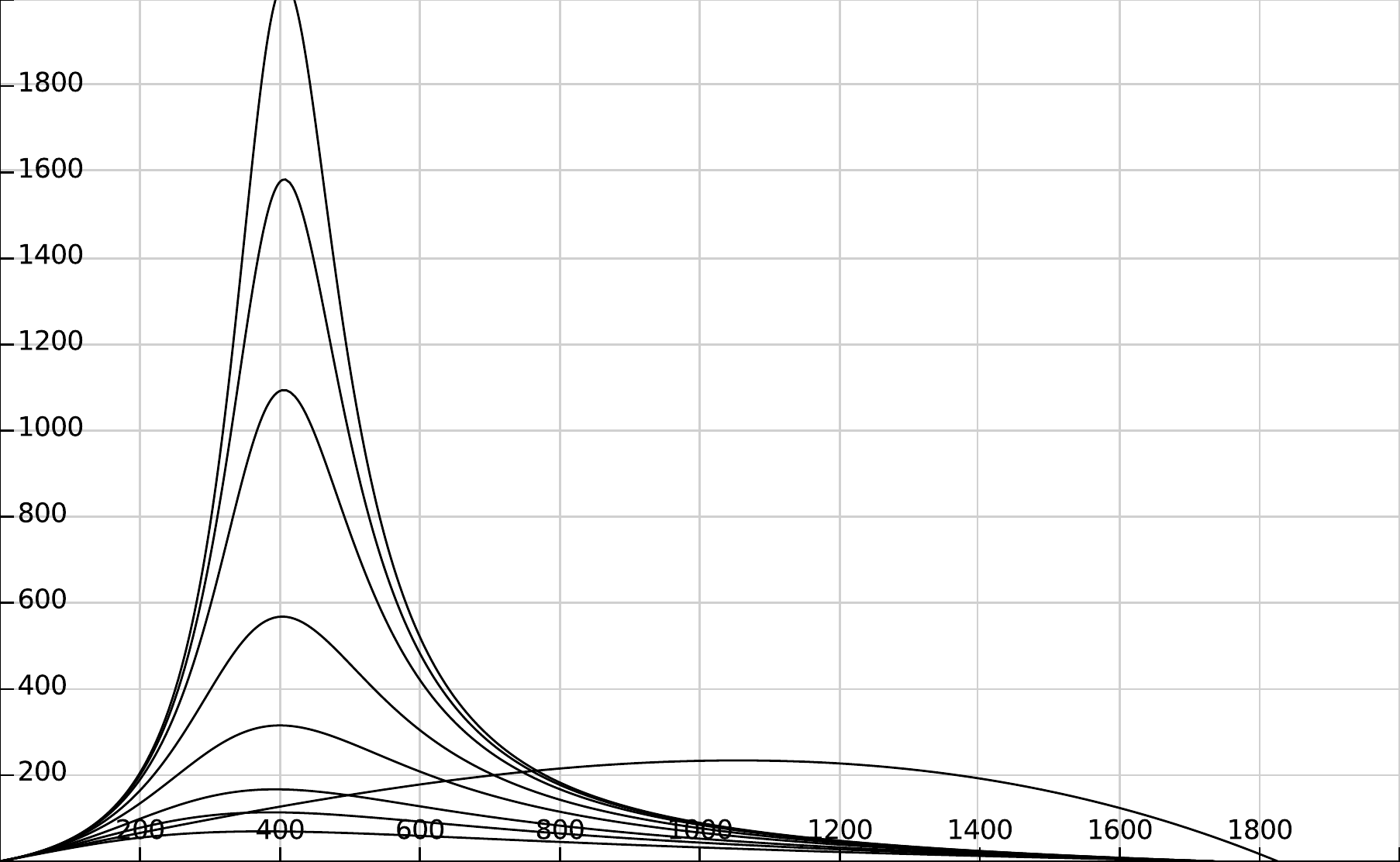}
\caption{Dispersion curves  \eqref{COMPAT-10}  and \eqref{COMPAT-11}  for 
$\mu_0=2, \gamma '  =0.1, \epsilon=0.006, C_1=D_1=0 $}
\vspace{0.1cm}
$ C_2-D_2=  0.1 \, (lower), 0.3, 0.4, 0.5, 0.55, 0.58, 0.58, 0.595 \, (upper) $
\label{F2}
\end{figure}

\vspace{0.5cm}

\begin{figure}[!h]
\centering
\vskip 0.2truecm
\includegraphics[scale=0.4]{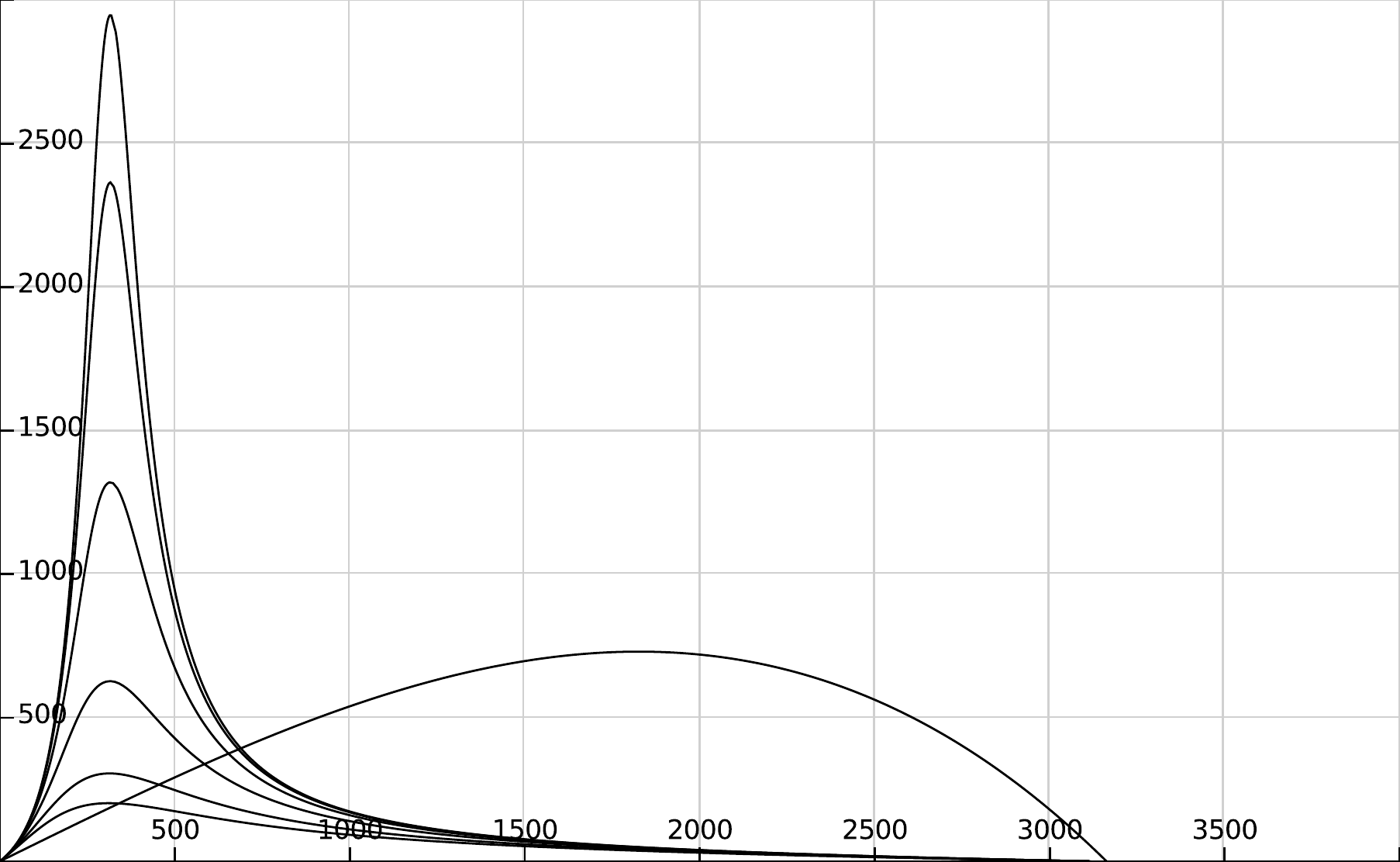}
\caption{Dispersion curves  \eqref{COMPAT-10}  and \eqref{COMPAT-11}  for 
$\mu_0=4, \gamma '  =0.1, \epsilon=0.006, C_1=D_1=0 $}
\vspace{0.1cm}
$ C_2-D_2=  0.1 \,  (lower), 0.2, 0.3, 0.35, 0.37, 0.375  \, (upper) $
\label{F3}
\end{figure}

\newpage

\begin{figure}[!h]
\centering
\vskip 0.2truecm
\includegraphics[scale=0.4]{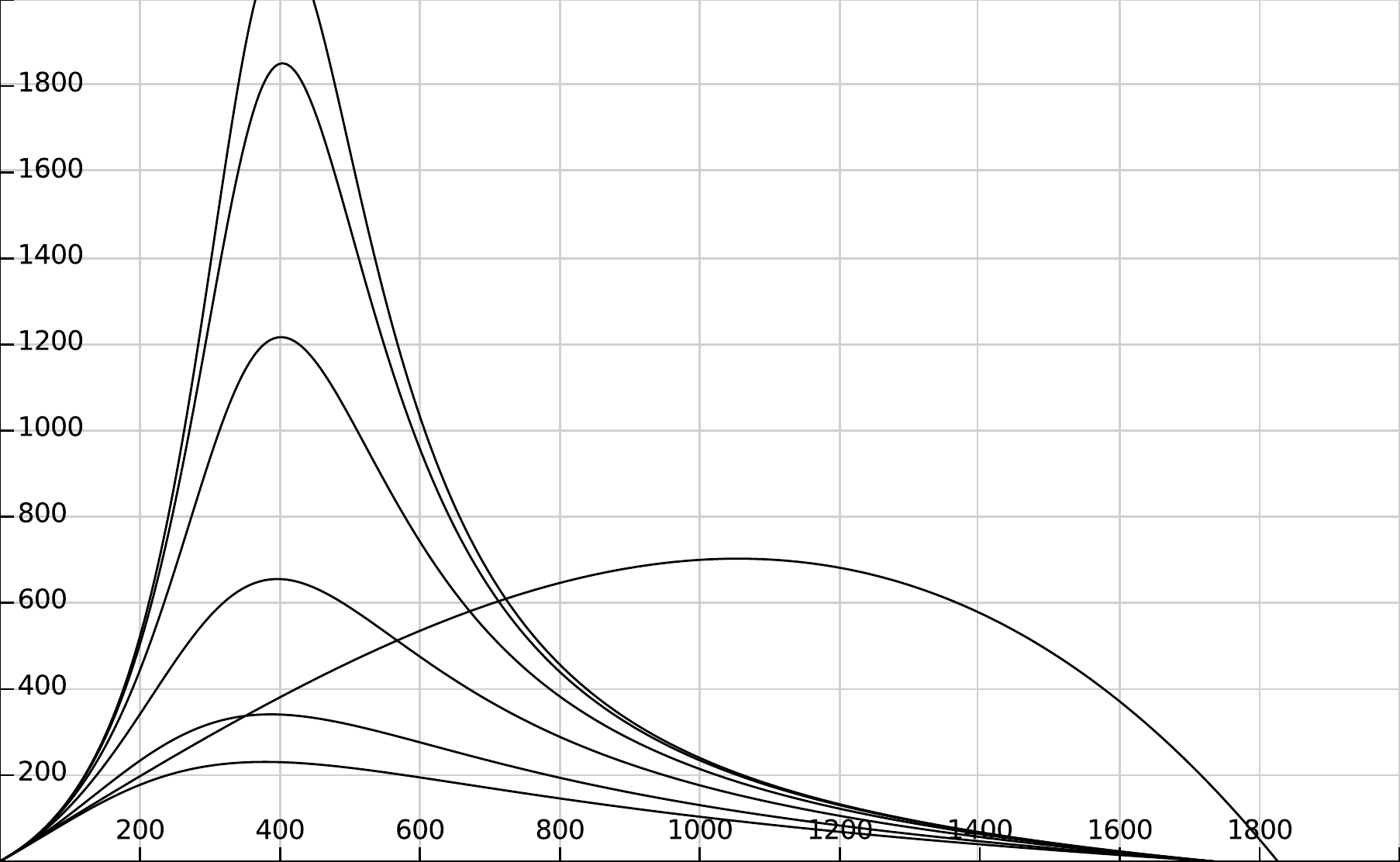}
\caption{Dispersion curves  \eqref{COMPAT-10A}  and \eqref{COMPAT-10B}  for 
$\gamma '  =0.1, \epsilon=0.006, C_1=D_1= 0 $}
\vspace{0.1cm}
$ C_2-D_2=   0.1  \, (lower), 0.2, 0.3, 0.35, 0.37, 0.375 \, (upper) $
\label{F4}
\end{figure}

\vspace{0.5cm}

\begin{figure}[!h]
\centering
\vskip 0.2truecm
\includegraphics[scale=0.4]{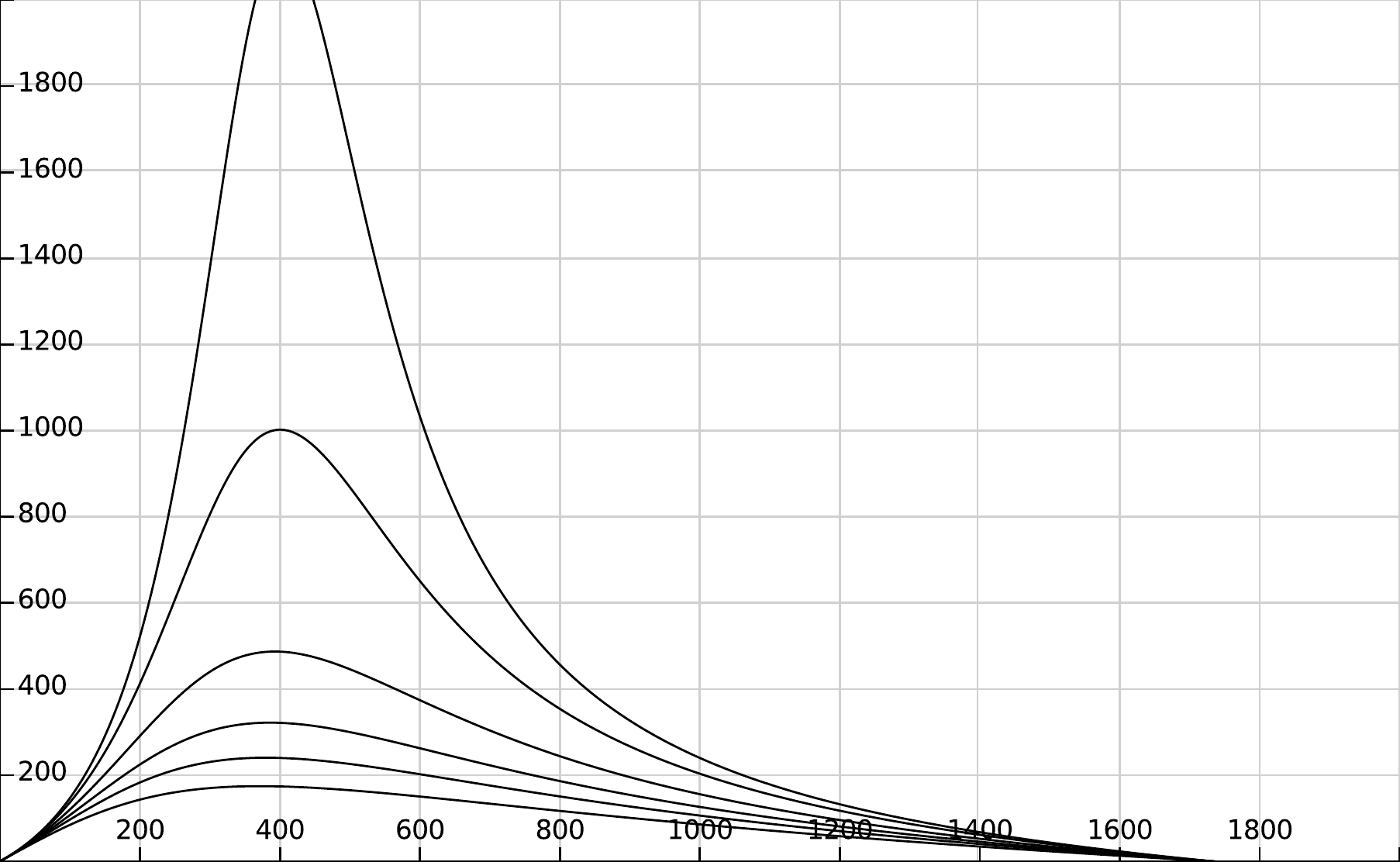}
\caption{Dispersion curves  \eqref{COMPAT-10A}   for 
$\gamma '  =0.1, \epsilon=0.006, C_2=0.375, r=D_2/C_2 $}
\vspace{0.1cm}
$ r=  1  \, (lower), 0.7, 0.5, 0.3, 0.1, 0  \, (upper) $
\label{F5}
\end{figure}


\section{ Conclusions}\label{CONC} 

{

In the last decades, some important results were established  concerning the 
linear stability of the displacement of  immiscible  non-Newtonian  fluids 
in Hele-Shaw cells. 

The displacement of a power-law fluid by air in a rectilinear cell was
studied by  Wilson ~\cite{WI} and   a formula of 
the growth  rate of perturbations was given, but  there is no 
qualitative change compared with the Saffman-Taylor result. The case of  radial 
displacements is different and  was studied in subsequent papers; the effect of 
the interfacial tension was highlighted.

Numerical results were obtained concerning the displacement of an Oldroyd-B
fluid  by air in rectilinear cells. A blow-up of the numerical growth rate was
reported, in accord with some experimental results concerning the flow of complex
fluids in Hele-Shaw cells.  On the other hand, most numerical methods shows the 
existence of a critical value of the Weissenberg numbers beyond which no discrete 
solutions can be obtained. 

In this paper we study  the linear instability  of the steady  displacement of two  
Oldroyd-B fluids  in a rectilinear Hele-Shaw cell.
{ We use the  Fourier decomposition \eqref{SA2A} for the velocities
$(u,v)$  and obtain the  formula} \eqref{COMPAT-10} of the growth rate of disturbances, 
which presents a blow-up for some critical values  of the Weissenberg  numbers. 

In the case of two Newtonian displacing  fluids, our growth rate is less than the 
Saffman-Taylor  value, but no  qualitative change appears - see Figure 1.
 We prove that the flow  stability is decided not only by the ratio of the 
displacing fluids viscosities, but also by the surface tension on the interface - 
see the  relations  \eqref{FINAL003A} - \eqref{FINAL003C}.
The Saffman - Taylor viscous fingering problem in rectangular geometry is studied 
in \cite{MI}, highlighting the link between interface asymmetry and viscosity contrast. 
The equation (19) of \cite{MI} shows that the growth rates will not become positive 
if the surface tension is large enough. This is in agreement with our result A3) in 
section 7. 

In the case of  two Oldroyd-B displacing fluids we get the instability criterion 
\eqref{COMPAT-14}. A strong destabilization effect  appears, compared with the Newtonian 
displacements.
The dispersion curves when the displacing fluid is Stokes (or air) are plotted 
in  Figures 2-5. 
Our analytical results are quite similar with numerical results already obtained 
{ in \cite{WI}}, then   the instability is due to the flow model,  at least 
for the  flow geometry   considered here.

\vspace{1cm}

{
{\bf Appendix 1 - the equations of the basic extra-stress tensors  $\tau^{0i}$. } 
}

The basic extra-stress tensors in both fluids $i=1,2$ are obtained from \eqref{A3} -
\eqref{A9}:
$$
{\bf \tau}^{0i} -  c_i ( {\bf V}^{0i} {\bf \tau}^{0i} + {\bf \tau}^{0i} {\bf V}^{0iT}) = $$
\begin{equation}\label{A11}
2 \mu_i \{ {\bf S}^{0i}  - d_i ({\bf V}^{0i} {\bf S}^{0i} + {\bf S}^{0i} {\bf V}^{0iT})\}.
\end{equation}

$$ ({\bf V}^{0i} {\bf \tau}^{0i} + {\bf \tau}^{0i} {\bf V}^{0iT})_{11} =  2 u^{0i}_z {\bf \tau}^{0i}_{31},  $$
$$ ({\bf V}^{0i} {\bf \tau}^{0i} + {\bf \tau}^{0i} {\bf V}^{0iT})_{12} =  u^0_z {\bf \tau}^{0i}_{32},       $$
$$ ({\bf V}^{0i} {\bf \tau}^{0i} + {\bf \tau}^{0i} {\bf V}^{0iT})_{13} =    u^{0i}_z {\bf \tau}^{0i}_{33},  $$
$$ ({\bf V}^{0i} {\bf \tau}^{0i} + {\bf \tau}^{0i} {\bf V}^{0iT})_{22} = 0, $$
$$ ({\bf V}^{0i} {\bf \tau}^{0i} + {\bf \tau}^{0i} {\bf V}^{0iT})_{23} = 0, $$
\begin{equation}\label{A11AB} 
({\bf V}^{0i} {\bf \tau}^{0i} + {\bf \tau}^{0i} {\bf V}^{0iT})_{33}  =  0. 
\end{equation}

$$  2( {\bf V}^{0i} {\bf S}^{0i} + {\bf S}^{0i} {\bf V}^{0iT}) _{11}   =    2(u^{0i}_z)^2,          $$                  
\begin{equation}\label{A11AC} 
  2( {\bf V}^{0i} {\bf S}^{0i} + {\bf S}^{0i} {\bf V}^{0iT}) _{ij}     = 0, \,\, (ij)\neq (11).   
\end{equation}   

\vspace{0.5cm}                                                                                    
{
{\bf Appendix 2 - the tensors E, F, S  in the formula \eqref{P4}. }
}
$$
{\bf E}: = {\bf V}^0 {\bf \tau} + {\bf V} {\bf \tau}^0 + {\bf \tau}^0 {\bf V}^T + 
{\bf \tau} {\bf V}^{0T},                                                                      $$
$$ {\bf E}_{11} =  2(u^0_z {\bf \tau}_{31}+ u_x {\bf \tau}^0_{11} + {\bf \tau}^0_{13}u_z),    $$
$$ {\bf E}_{12} =  (u^0_z {\bf \tau}_{32}+ v_x {\bf \tau}^0_{11}+ v_z {\bf \tau}^0_{13}),     $$
$$ {\bf E}_{13} =  (u^0_z {\bf \tau}_{33} +   u_x {\bf \tau}^0_{13}),                         $$
\begin{equation}\label{P6}
 {\bf E}_{22} = 0,  \quad   {\bf E}_{23} =  {\bf \tau}^0_{13} v_x , \quad   {\bf E}_{33}= 0.                                                                                          
\end{equation}

$$ {\bf F}: = 2[ {\bf V}^0 {\bf S} + ({\bf V}^0 {\bf S})^T + {\bf V} {\bf S}^0 + 
({\bf V} {\bf S}^0)^T ],                                                                 $$ 
$$  {\bf F}_{11} = 4u^0_zu_z, \,\,    {\bf F}_{12} = 2u^0_zv_z,                          $$
$$  {\bf F}_{13} = u_xu^0_z,                                                             $$
\begin{equation}\label{P6A}
  {\bf F}_{22} = {\bf F}_{33}=0, \,\,  {\bf F}_{23}  =  v_xu^0_z .  
\end{equation}

$$ 2 {\bf S}_{11}= 2u_{x}, \,\,   2 {\bf S}_{12} = (u_y+v_x),        $$
$$ 2 {\bf S}_{13} = u_z,    \,\    2 {\bf S}_{22} =  v_y,            $$
\begin{equation}\label{P7}  
2 {\bf S}_{23} = v_z,   \,\     2 {\bf S}_{33} =  0.                
\end{equation}
 
\vspace{1cm}

{
{\bf Appendix 3 - the extra-stress  perturbations $\tau^i$. }
}

1) We use $D_i, C_i$ given in  \eqref{COMPAT-9}. From  \eqref{P4} we  get the dimensionless 
constitutive    relations 
for 	  $\tau^1_{33},  \tau^1_{32},  \tau^1_{31}$:																																							
\begin{equation}\label{PRIMELE-A} 
\tau^1_{33} + C_1  \epsilon u^0 \tau^1_{33,x} = 0,                                  
\end{equation}
$$
   \tau^1_{32} + C_1 \epsilon  (u^0\tau^1_{32,x} -  \tau^{0}_{13}v^1_x ) =      $$
\begin{equation}\label{PRIMELE-2}
v^1_z  + D_1  \epsilon (u^{0}v^1_{zx} - v^1_x u^{0}_z)   .                        
 \end{equation}               
$$
 \tau^1_{31}  +C_1  \epsilon u^0\tau^1_{31,x} -
C_1  \epsilon (u^0_z \tau^1_{33} + u^1_x\tau^{0}_{13})=                         $$
\begin{equation}\label{PRIMELE-1}
 u^1_z +  D_1  \epsilon( u^{0}u^1_{zx} - u^1_xu^{0}_z).                                        
\end{equation}             
At the leading order, from \eqref{PRIMELE-A} we get   
\begin{equation}\label{PRIMELE-3}
\tau^1_{33}=0.
\end{equation}  
The partial derivatives $u_x, u_y, v_x, v_y$ near $x=0$ are bounded in  terms of $k$, due 
to the Fourier decomposition  \eqref{SA1}   - \eqref{SA3}  with  $\alpha \neq 0$. 
{  Moreover, we have}  
\begin{equation}\label{PRIMELE-3A0}
(u^{0}v^1_{zx} - v^1_x u^{0}_z)=0,  \,\, (u^{0}u^1_{zx} - u^1_x u^{0}_z)=0,         
\end{equation}                            
then from the  relations  \eqref{PRIMELE-2} - \eqref{PRIMELE-3A0}  it follows 
$$
   \tau^1_{32} + C_1 \epsilon  (u^0\tau^1_{32,x} -  \tau^{0}_{13}v^1_x ) =  v^1_z, $$
 \begin{equation}\label{PRIMELE-3A}
   \tau^1_{31} + C_1 \epsilon  (u^0\tau^1_{31,x} -  \tau^{0}_{13}u^1_x ) = 
                                                                 {  u^1_z}.
\end{equation}
{The dimensionless form of $\eqref{A12}_1$ is $ \tau^{0i}_{13} = u^{0i}_z $,  
therefore  from the last relations it follows} 
$$
   \tau^1_{32} + C_1 \epsilon  (u^0\tau^1_{32,x} -  u^{0}_zv^1_x ) =  v^1_z,         $$
 \begin{equation}\label{PRIMELE-3AB}
  \tau^1_{31} + C_1 \epsilon  (u^0\tau^1_{31,x} -  u^{0}_zu^1_x ) =   
                                                                    { u^1_z}.
\end{equation}
{ We use again \eqref{PRIMELE-3A0} and relations \eqref{PRIMELE-3AB} lead us 
to}   
\begin{equation}\label{PRIMELE-4}
\tau^1_{32}=v^1_z   , \,\,  \tau^1_{31}=u^1_z.
\end{equation} 
In  \eqref{PRIMELE-A} - \eqref{PRIMELE-4} replace  $C_1, D_1$ by $C_2, D_2$,  then for 
$i=1,2$ we obtain 
\begin{equation}\label{PRIMELE-5}
\tau^i_{33}=0,  \, \tau^i_{31}=u^i_z   , \,\,  \tau^i_{32}=v^i_z.
\end{equation}

2)  The dimensionless form of the relations  \eqref{P4} gives us
the component of $\tau_{11,x}$ in the fluid 1 (in the formulas \eqref{TAU-11-A} -
\eqref{TAU-11-D} below we omit the upper index $^1$):
$$ \tau_{11,x} + C_1 u^0 \tau_{11,xx} \epsilon  -                                   $$
$$ 2 C_1[ u^0_z u_{zx} / \epsilon  +  
 u_{xx}2 (C_1-D_1)(u^0_z)^2  +   u^0_zu_{zx} / \epsilon ]  =                        $$ 
\begin{equation}\label{TAU-11-A}
  2u_{xx}  + 2 D_1 u^0 u_{xxx}  \epsilon  -    4 D_1 u^0_zu_{zx} / \epsilon.     
\end{equation}
We suppose 
\begin{equation}\label{TAU-11-B}
 \tau_{11x} =   2u_{xx}  + 4  (C_1-D_1) u^0_zu_{zx}/ \epsilon   . 
\end{equation}
We insert the expression \eqref{TAU-11-B} in the equation \eqref{TAU-11-A}  
and get 
$$
\tau_{11,x} =    2u_{xx} +   4(C_1-D_1) u^{0}_{z}u_{zx} / \epsilon+                $$     
$$  2 (D_1 - C_1) \epsilon u^{0}  u_{xxx}  +                                       $$
\begin{equation}\label{TAU-11-C}			
4C_1(C_1-D_1)[u_{xx}(u^0_z)^2 -u^0 u^0_zu_{zxx}].
\end{equation}
The Fourier decomposition \eqref{SA1} gives us 
$$ u_{xx}(u^0_z)^2 -u^0u^0_zu_{zxx} =0.                                            $$ 
Then the formula \eqref{TAU-11-B} { is verified}  if we can neglect the term  
$2 (C_1 - D_1) \epsilon u^{0}  u_{xxx}$  in front of the first two terms appearing in 
the equation  \eqref{TAU-11-C}.  We recall that $u_x, u_{xx}, u_{xxx}$  are  bounded
{ with respect to }  $k$, then 
$$  u_{xx}= O(\epsilon^2),                                                   $$
$$   4(C_1-D_1) u^{0}_{z}u_{zx} / \epsilon = O(\epsilon),                   $$
 \begin{equation}\label{TAU-11-D}  
	2 (D_1 - C_1) \epsilon u^{0}  u_{xxx}  = O(\epsilon^3).                       
\end{equation}	
We neglect the term of order $O(\epsilon^3)$ in  the equation  \eqref{TAU-11-C} 
and obtain the approximate formula 
\begin{equation}\label{TAU-11-E}
 \tau^1_{11x} =   2u^1_{xx}  + 4  (C_1-D_1) u^{0}_zu^1_{zx}/ \epsilon ,
\end{equation}
in agreement with the hypothesis \eqref{TAU-11-B}.

On the same way we get the approximate formula of $\tau_{11,x}$ in the second
fluid:
\begin{equation}\label{TAU-11-F}
 \tau^2_{11x} =   2u^2_{xx}  + 4  (C_2-D_2) u^{0}_zu^2_{zx}/ \epsilon   . 
\end{equation}
{\it We emphasize that  $u_{xx}, u_{yy}$ { appear}  in the approximate curvature
of $\psi$, so they can not be neglected}.

\vspace{0.25cm}

3)  The dimensionless form of the relations  \eqref{P4}  gives us
the component of $\tau_{12,y}$ in the fluid 1 (in the formulas 
\eqref{TAU-12-A} - \eqref{TAU-12-D} below we omit the index $^1$):  
$$ 
\tau_{12,y} + C_1 u^0\tau_{12,xy} \epsilon -                                                    $$
$$  C_1 [ 2 u^0_z v_{zy}/ \epsilon + 2 (C_1-D_1) v_{zy} (u^0_z)^2  ]  =             $$
\begin{equation}\label{TAU-12-A}   
(u_y+v_x)_y +   D_1  u^0(u_y+v_x)_{xy} \epsilon    -  2D_1 u^0_zv_{zy} / \epsilon .     
\end{equation}
We suppose 
\begin{equation}\label{TAU-12-B} 
  \tau_{12,y}  =  
  (u_y+v_x)_y  + 2  (C_1 -D_1) u^0_{z} v_{zy}/ \epsilon    .   
\end{equation}
We introduce the expression  \eqref{TAU-12-B} in  \eqref{TAU-12-A}  and   get 
$$  \tau_{12,y}=    (u_y + v_x)_y +  2(C_1-D_1) u^0_z v_{zy}/ \epsilon +         $$
$$   (D_1-C_1)   u^0(u_y+v_x)_{xy} \epsilon   +                                              $$                   
\begin{equation}\label{TAU-12-C}  
2 C_1 (C_1-D_1)[  v_{xy} (u^0_z)^2  - u^0u^0_z  v_{zyx} ] .                          
\end{equation}
The Fourier decomposition \eqref{SA1} gives us  
$$v_{xy}(u^0_z)^2 -u^0u^0_zv_{zxy} =0.                                                       $$  
If $C_i, D_i =O(1)$ we get 
$$   (u_y + v_x)_y =O(\epsilon^2),                                                            $$  
$$  2(C_1-D_1) u^0_z v_{zy}/ \epsilon =  O(\epsilon),                                $$
\begin{equation}\label{TAU-12-D}
   (D_1-C_1)   u^0(u_y+v_x)_{xy} \epsilon  =O(\epsilon^3).
\end{equation} 
We neglect the term of order $O(\epsilon^3)$ in  the equation  \eqref{TAU-12-C} 
and  get  the formula \eqref{TAU-12-B}.   Therefore, for $i=1,2$,  we have 
\begin{equation}\label{TAU-12-F}
\tau^i_{12,y} =  (u^i_y + v^i_x)_y +  2(C_i-D_i) u^{0}_z v^i_{zy}/ \epsilon  . 
\end{equation}

4) The dimensionless constitutive relation  for   $\tau^1_{22}$   is  
\begin{equation}\label{TAU-22-A}
 \tau^1_{22} + C_1 \epsilon u^{0} \tau^1_{22,x}  = 2 v^1_{y} + 
D_1  \epsilon u^{0} v_{yx} 
\end{equation}
and  at the leading order  we get 
\begin{equation}\label{TAU-22-B}
 \tau^i_{22}  = 2 v^i_{y}, \quad i=1,2.
\end{equation}


\end{document}